 \documentclass{emulateapj}
\usepackage{subfigure}
\usepackage{verbatim}

\slugcomment{To appear in Astrophysical Journal}
\date{Received December 13, 2011; accepted March 5, 2012}
\shorttitle{SOLAR RADIUS FROM THE 2003 AND 2006 MERCURY TRANSIT}
\shortauthors{Emilio et al.}

\begin{document}

\title{MEASURING THE SOLAR RADIUS FROM SPACE DURING THE 2003 and 2006 MERCURY TRANSITS}

\author{M. Emilio\altaffilmark{1}}
\author{J. R. Kuhn\altaffilmark{2}}
\author{R. I. Bush\altaffilmark{3}}
\author{I. F. Scholl\altaffilmark{2}}

\altaffiltext{1}{Observat\'orio Astron\^omico-Departamento de
Geoci\^encias Universidade Estadual de Ponta Grossa, Paran\'a,
Brazil, memilio@uepg.br}

\altaffiltext{2}{Institute for Astronomy, University of
Hawaii, 2680 Woodlawn Dr. 96822, HI, USA, kuhn@ifa.hawaii.edu, ifscholl@hawaii.edu}

\altaffiltext{3}{HEPL-Stanford University, Stanford, CA, 94305, USA, rock@sun.stanford.edu}

\begin{abstract}
The Michelson Doppler Imager (MDI) aboard the \emph{Solar and
Heliospheric Observatory} observed the transits of Mercury on 2003 May 7 and 2006 November 8.  Contact times between Mercury and the solar limb have been used since the 17th century
to derive the Sun's size but this is the first time that high-quality imagery from space, above the Earth's
atmosphere, has been available.  Unlike other measurements this technique is
largely independent of optical distortion.
The true solar radius is still a matter of debate in the literature as measured differences of several tenths of an
arcsecond (i.e., about 500 km) are apparent. This is due mainly to systematic
errors from different instruments and observers since the claimed
uncertainties for a single instrument are typically an order of magnitude
smaller. From the MDI transit data we find the solar radius to be $960".12
\pm 0".09 $ $(696,342 \pm 65 \mathrm{km} )$. This value is consistent between the transits and consistent between
different MDI focus settings after accounting for systematic effects.
\end{abstract}


\section{Introduction}

Observations of the interval of time that the planet Mercury takes to transit in front of the Sun provides, in principle, one of the most accurate methods to measure the solar diameter and potentially its long-term variation. Ground observations are limited by the spatial resolution with which one can determine the instant Mercury crosses the limb. Atmospheric seeing and the intensity gradient near the limb (sometimes called the ``black drop effect'', see \citet{Schneider04}; \citet{pas05}) contribute as error sources for the precise timing required to derive
an accurate radius. This is the first time accurate Mercury's transit contact times is measured by an instrument in space and improves at least 10 times the accuracy of classical observations (see \citet{bes32}; \citet{gam32}).

About 2,400 observations of those contacts from 30 transits of Mercury, distributed during the last 250 years, were published by \citet{morrison75}, and analyzed by \citet{parkinson80}. Those measurements, collected mainly to determine the variations of rotation of the Earth and the relativistic movement of Mercury's perihelion provided a time-series of the
Sun's diameter. Analyzing this data set, \citet{parkinson80} marginally found a decrease in the solar semi-diameter of $0".14 \pm 0".08$ from 1723 to 1973, consistent with the analysis of \citet{shapiro80} which claimed a decrease of 0".15 per century. Those variations are consistent with our \citep{bush10} null result and upper limit to secular variations obtained from Michelson Doppler Imager (MDI) imagery of 0".12 per century. \citet{Sveshnikov02}, analyzing 4500 archival contact-timings between 1631 and 1973, found that the secular decrease did not exceed $0".06 \pm 0".03$.

\begin{figure*}
 \centering
 \plotone{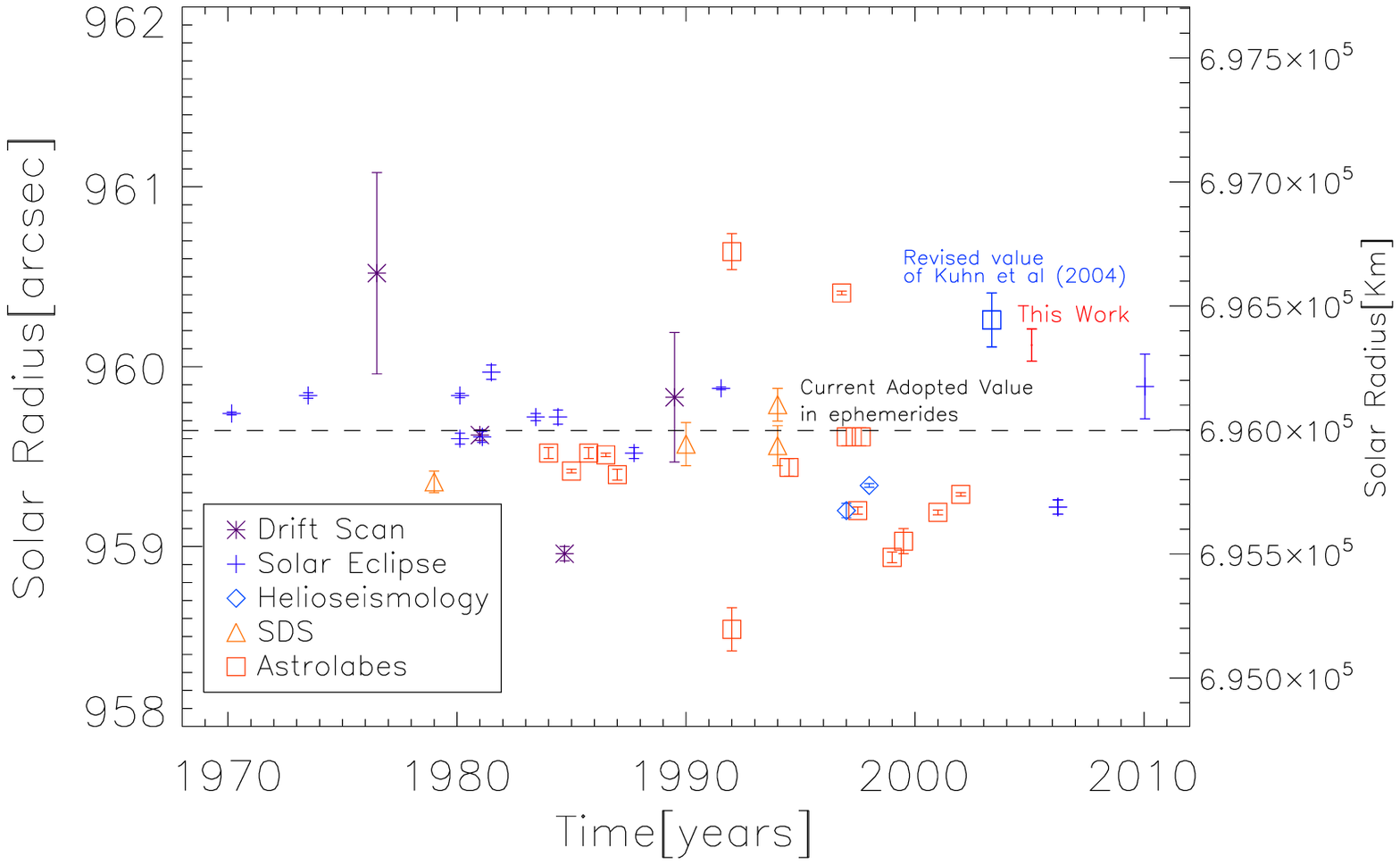}
 \caption{Published measurements of solar radius in the last 32 years. Abscissa values are the mean observation dates. The plus signs are drift scan measurements and they are from \citet{wittmann80, yoshizawa94, neckel95, brown98}. The solar eclipses measurements are marked with an asterisk and are from \citet{fiala84, kubo93, akimov93, kilic09, sigismondi09, adassuriya11}. Helioseismology measurements (diamonds) have a different definition of solar radius and can differ up to 300 km from the inflection point definition \citep{Tripathy99}. Helioseismology data are from \citet{schou97, antia98}.  SDS measurements are show in triangle and are from \citet{sofia83, maier92, egidi06, djafer08}. After applying systematic corrections to SDS measurement \citet{djafer08} found the value of $959".798 \pm 0".091$ (showed in this figure as the upper triangle point in 1994). This value is inside the $2 \sigma$ error bar of this work. Published astrolabes measurements are show in square symbols and are from \citet{noel95, sanchez95, leiter96, laclare96, kilic98, jilinski99, laclare99, puliaev00, golbasi01, emilio01, andrei04, noel04, emilio05, kilic05}. The astrolabe measurements showed here are not correct for the atmospheric and instrumental systematic effects. When those corrections are made the final value is bigger than the ones displayed in this plot (see \citet{Chollet99}).
 \label{lradius}}
\end{figure*}

Modern values found in the literature for the solar radius range from $958".54 \pm 0".12$ \citep{sanchez95} to $960".62 \pm 0".02$ \citep{wittmann03}. Figure~\ref{lradius} shows published measurements of solar radius over the last 30 years (for a review see: \citet{kuhn04, emilio05, Thuillier05}). Evidently these uncertainties reflect the statistical errors from averaging many measurements by single instruments and not the systematic errors between measurement techniques. For example, our previous determination of the Sun's radius with MDI was based on an optical model of all instrumental distortion sources \citep{kuhn04}. The method described here has only a "second-order" dependence on optical distortion (although it is sensitive to other systematics) and may be more accurate for this reason.

 \begin{figure}
 \plotone{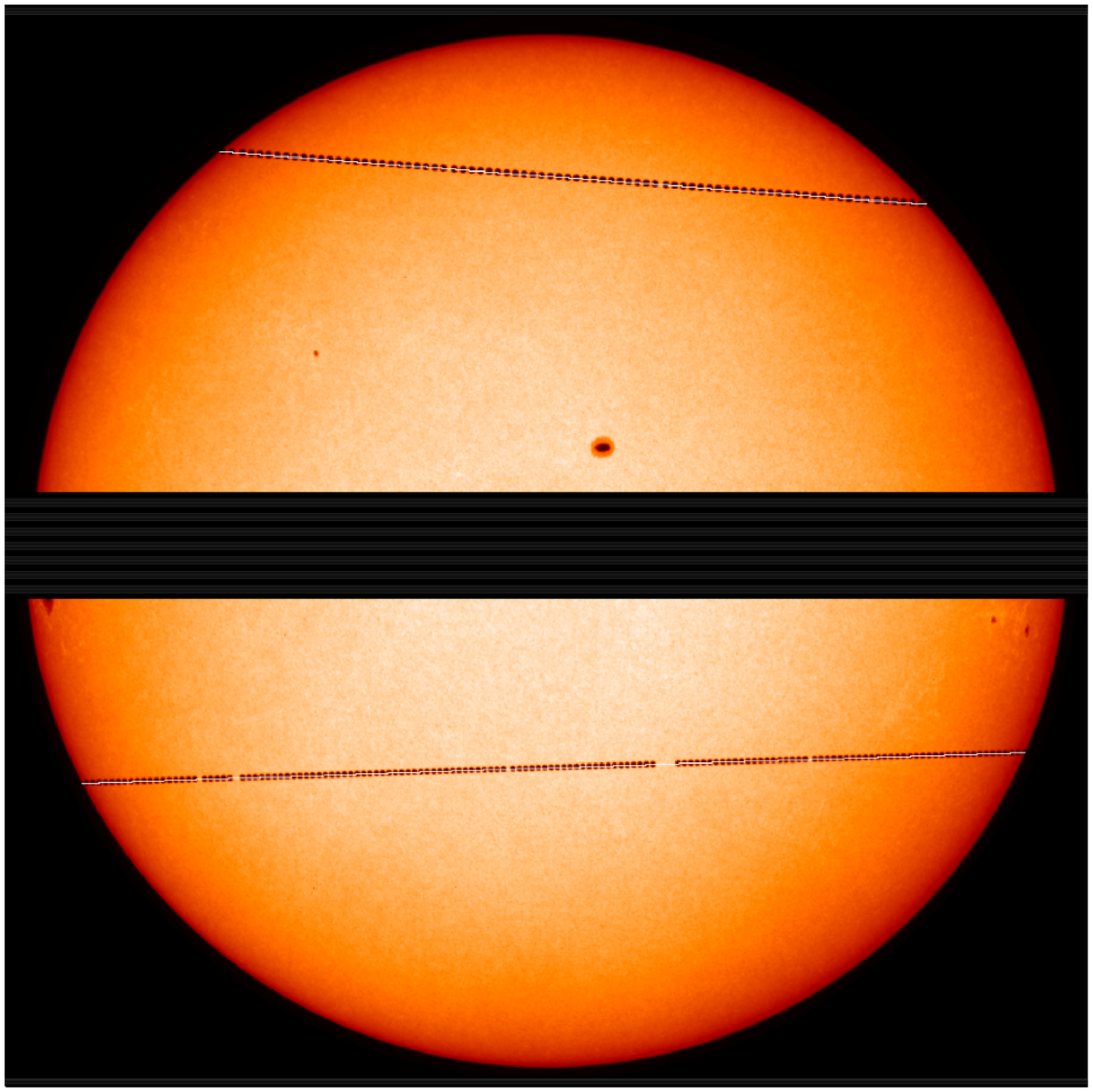}
 \caption{Composite image of two Mercury transits in
 2003 May (top, 4 minute cadence) and 2006 November (bottom, 2 minute
 cadence). Images were taken with one-minute cadence in both
 transits, intercalating four different focus blocks in 2003 and two focus
 blocks in 2006. Both images showed here are at focus settings 4.
 \label{transit}}
 \end{figure}

\section{Data Analysis}

The data consist of $1024\times1024$ pixel MDI-SOHO images from fixed wavelength
filtergrams of Mercury crossing in front of the Sun. Images were
obtained with a one-minute cadence in both transits, cycling between four instrument
focus settings ("focus blocks", FBs) in 2003 and two in 2006. For each image obtained at a given focus
block, we subtracted a previous image of the Sun without Mercury using the same focus
setting. This minimized the effects of the limb darkening gradient and allowed us
to find the center of Mercury (Figure~\ref{transit})
 more accurately. In a small portion of each image containing Mercury we fitted a negative Gaussian.
We adopt the center of the Gaussian as the center of Mercury. Images that were too close
to the solar limb were not fitted because the black drop effect inflates our center-determination errors. The
position of the center of the Sun and the limb were calculated as described in
\citet{emilio00}. A polynomial was fit to the $x-y$ pixel coordinates of Mercury's center
during the image time-series. This transit trajectory was extrapolated to find the
precise geometric intersection  with the limb by also iteratively accounting
for the time variable apparent change in the solar radius ($-0.221\pm 0.004$  mas minute$^{-1}$ in 2003
and $0.152 \pm 0.003$  mas minute$^{-1}$ in 2006) between images.  %(Figure~\ref{ldftime}).

\begin{figure*}
    \centering
    \subfigure[2003 Mercury Transit - $1^{st}$ and $2^{nd}$ contacts]
    {
        \includegraphics[scale=.645]{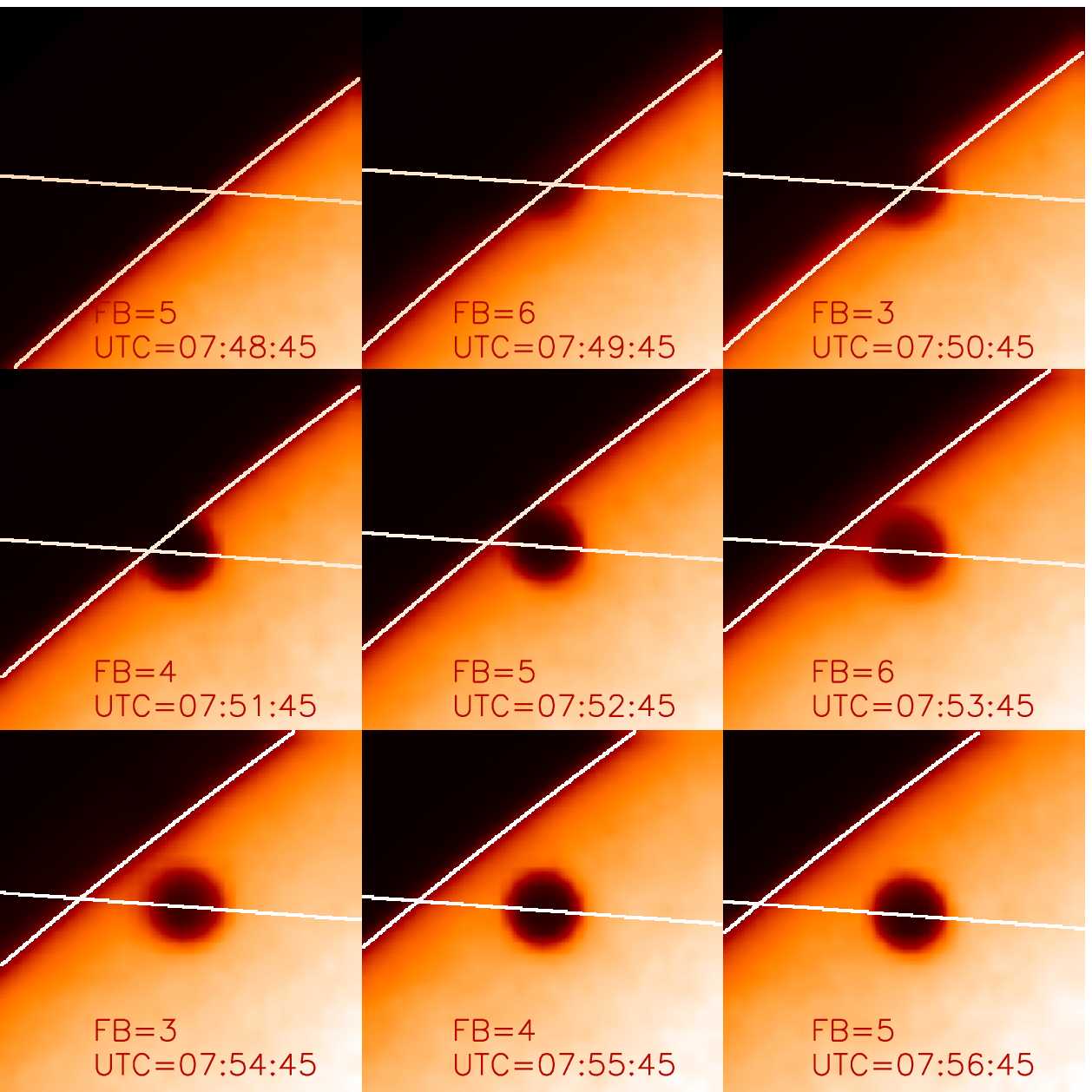}
        \label{fig:m2003a}
    }
     \subfigure[2003 Mercury Transit - $3^{rd}$ and $4^{th}$ contacts]
    {
        \includegraphics[scale=.645]{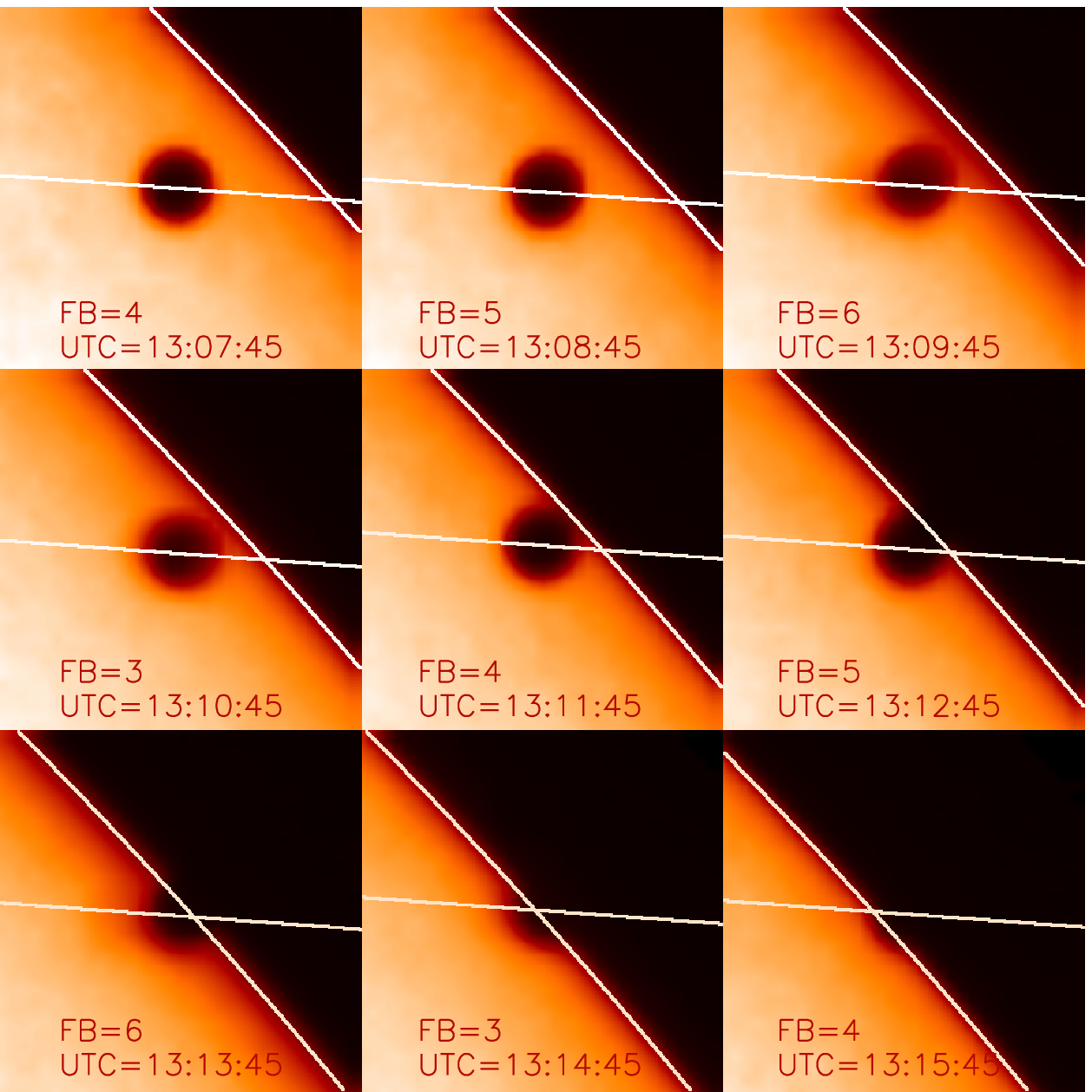}
        \label{fig:m2003b}
    }
    \\
    \subfigure[2006 Mercury Transit - $1^{st}$ and $2^{nd}$ contacts]
    {
        \includegraphics[scale=.645]{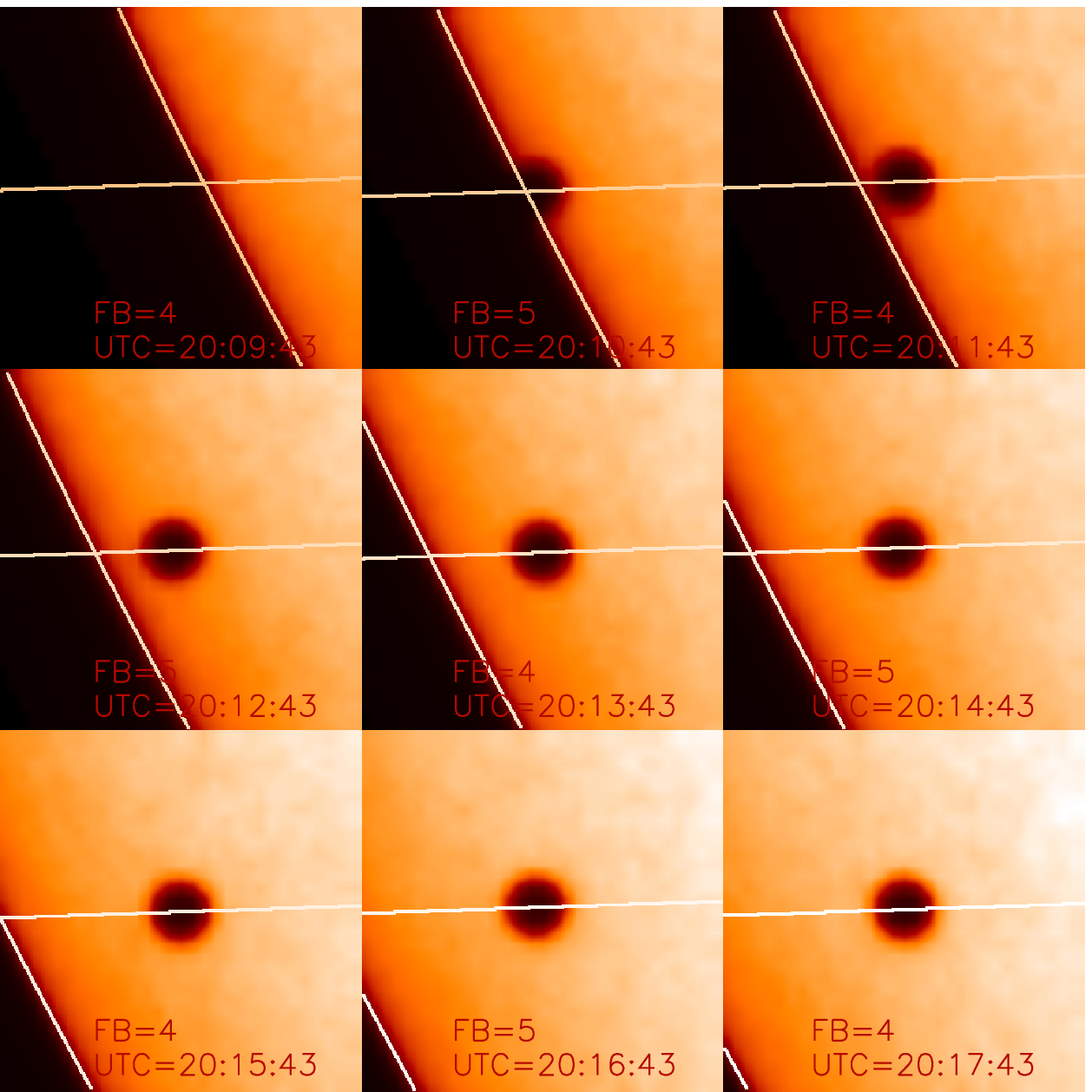}
        \label{fig:m2006a}
    }
    \subfigure[2006 Mercury Transit - $3^{rd}$ and $4^{th}$ contacts]
    {
        \includegraphics[scale=.645]{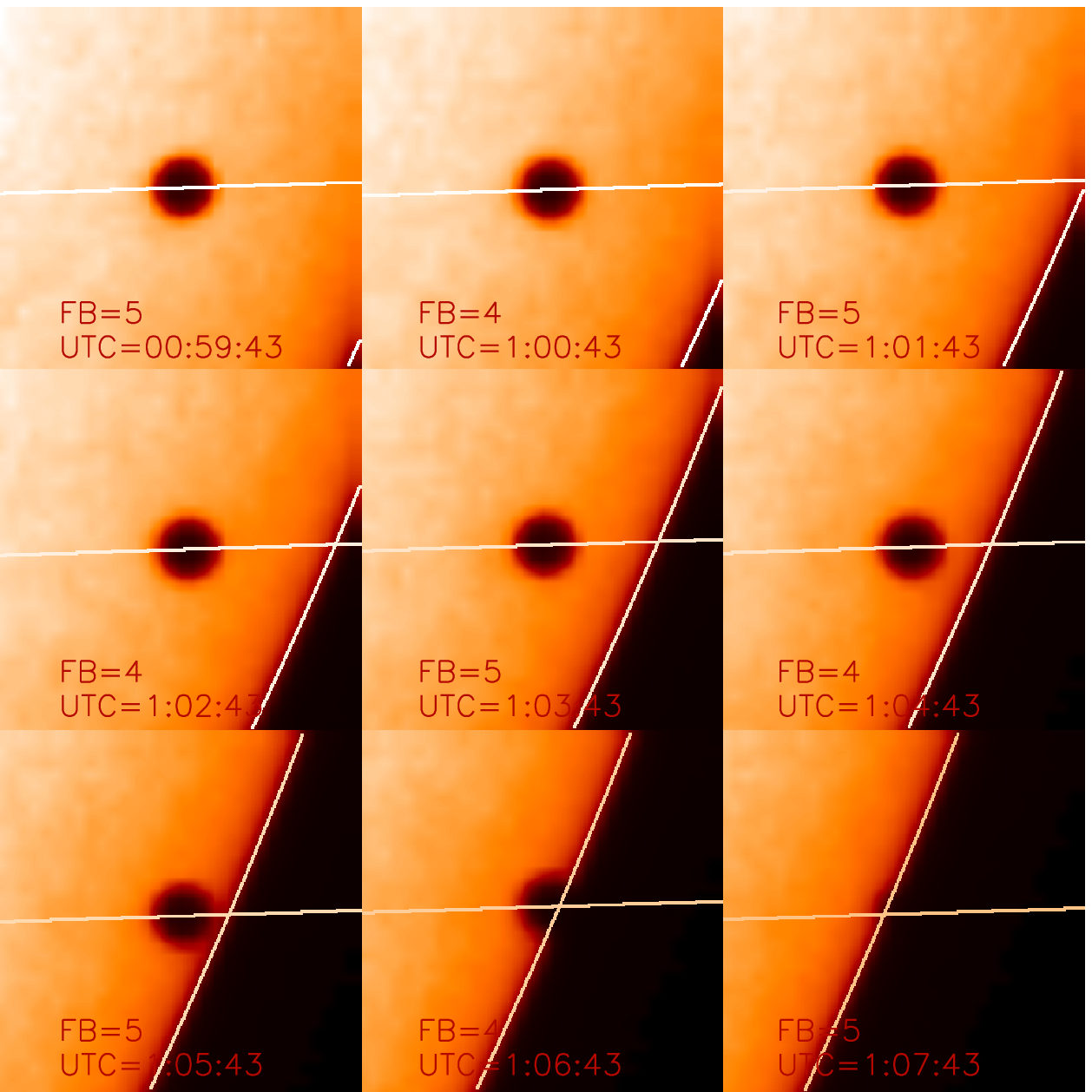}
        \label{fig:m2006b}
    }
    \caption{Images near the contact times for both 2003 and 2006 transits
     as well as the Mercury trajectory fit and the solar limb fit. Mercury center geometric x and y contact positions with the solar limb where found from the interception of those fits.(a) 2003 Mercury transit: first and second contacts, (b) 2003 Mercury transit: third and fourth contacts,(a) 2006 Mercury transit: first and second contacts, (b) 2006 Mercury transit: third and fourth contacts }
    \label{mercury_contacts}
\end{figure*}

\begin{deluxetable*}{lcccc}
\tabletypesize{\scriptsize}
\tablecaption{Mercury transit ephemerides}
\tablewidth{500pt}
\tablehead{\colhead{Year} & \colhead{First Mercury Contact \tablenotemark{a}} &
\colhead{Second Mercury Contact \tablenotemark{a}} & \colhead{Mercury Center Transit Duration} &
\colhead{Mercury Velocity} \\ \colhead{} & \colhead{} &
\colhead{} & \colhead{(minute)} & \colhead{(arcsec minute$^{-1}$)}}

\startdata
2003  &  50.433 \tablenotemark{b} &  374.342 \tablenotemark{b} &  323.909 & 4.102\\
2006  &  10.458 \tablenotemark{c} &  307.059 \tablenotemark{c} &  296.601 & 5.968\\
\enddata
\tablenotetext{a}{Center contacts in (min).}
\tablenotetext{b}{UTC-7 hr  of 2003 May 7.}
\tablenotetext{c}{UTC-20 hr of 2006 November 8.}
\label{ephemerides}
\end{deluxetable*}

\begin{figure*}
 \centering
 \plotone{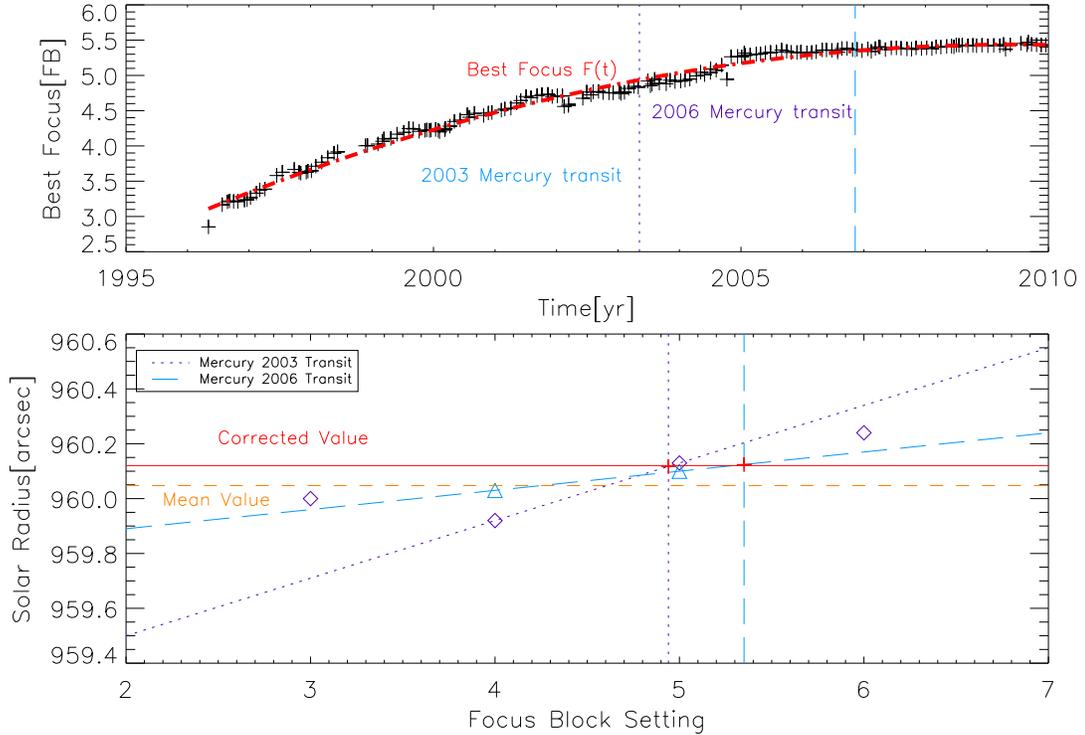}
 \caption{Top: the best focus position fit when the two mercury transits happened. 2003 Mercury transit date is shown as a dotted line and 2006 Mercury transit as a long dashed line. From the intersection of those lines with the best focus position fit (dashed line) we found that the best focus positions are 4.94 and 5.35 for 2003 and 2006 Mercury transits respectively.
  Bottom: linear fit as a function of focus block. Once again the 2003 Mercury transit best focus position is shown as a dotted line and 2006 Mercury transit as a long dashed line. The mean value is shown as a dashed line. We looked for the intercept positions with the best estimate of the focus for 2003 and 2006 Mercury transits (solid line). The intercept values with the linear fit are the same for both 2003 and 2006 and represent our correction for the changes in focus.}
 \label{fbfit}
\end{figure*}

\begin{figure*}
 \centering
 \plotone{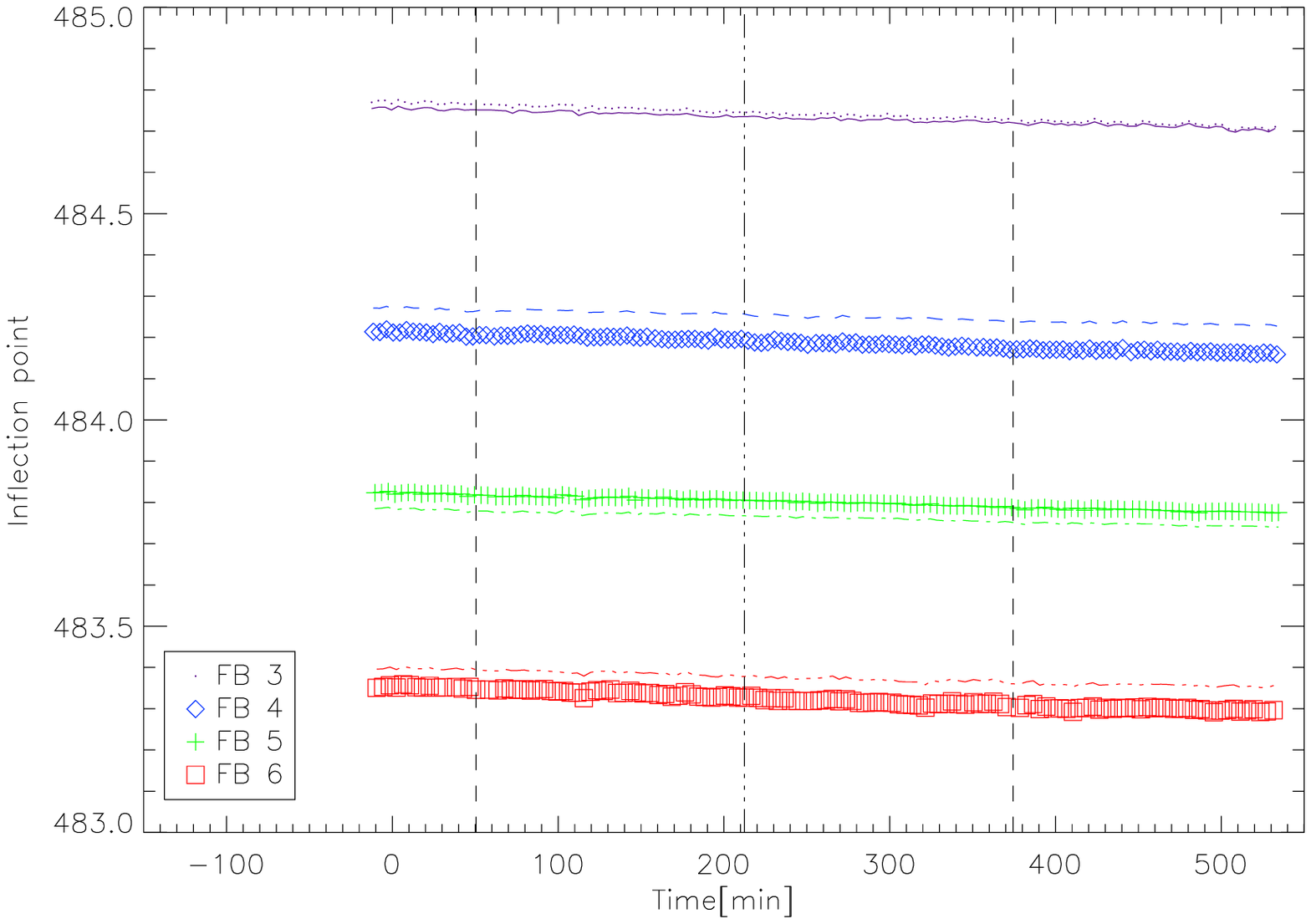}
 \caption{Different MDI focus positions result in a different size of the Sun in pixels. Both Mercury and Sun are subject to the same instrumental effects but since only the contact times are used to calculate the solar radius this effect is reduced. The linear trend of radius with time in this plot is due to the change of distance of SOHO from the Sun. All measurements were used to interpolate the solar size in pixels at the times of contacts (dashed lines). The triple-dot-dashed  line indicates the middle of 2003 Mercury transit. The figure also shows two different techniques to calculate the inflection points. The first one (dots) is a fit of a Gaussian plus quadratic to the squared radial derivative of the LDF. The second (lines) is fitting a parabola on the same function. Those differences reflect the systematic uncertainty in defining the solar radius.}
 \label{ldf_time}
\end{figure*}

The same procedure was performed independently for each FB time-series. The
contact times were found from a least-squares fit of Mercury's
center position trajectory $x^{FB}(t)$ and $y^{FB}(t)$ to the intersection with the limb near first/second and third/fourth contact.
Here FB are the FBs $(3,4,5,6)$. As a consequence we
obtained eight different contact times for 2003 and four for 2006,
one for each FB and two for each geometric contact. Figure~\ref{mercury_contacts}
shows zoom images of the Sun containing Mercury close to the contact times
for both 2003 and 2006 transits as well as the Mercury trajectory and the limb fit.

FB 3 and 6 are far from an ideal focus, as Figure~\ref{fbfit} (upper) and Figure~\ref{mercuryfb} show, so that the shape of the limb
darkening function (LDF) is different enough that it is difficult to compare corresponding "transit" points
with the near-focus images. Those focus block were not utilized in further analysis. We note also that FB 6 data contained a ghost image
of Mercury which added new systematic errors to the data.

The total transit time was thus obtained with an accuracy of 4 s in 2003 and 1 s in 2006 from a
comparison of FB 4 and FB 5 data. The correction to the radius is found from the
relation \citep{shapiro80}.
\begin{center}
$\Delta R=\frac{\omega^2 }{R_ \odot} T \Delta T^{O-C}$ \\
\end{center}
where $\omega$ is the speed of Mercury relative to the Sun, $T$ is
the total length of the transit, $R_\odot$ is the apparent value of
the solar radius at 1 AU from the ephemeris for each transit
instant (where we adopt 959".645 $(696,000 \pm 40 \mathrm{km})$ for the nominal solar radius) and
$\Delta T^{O-C}$ is the difference between the observed and ephemeris duration of the transit.
Table~\ref{ephemerides} shows the ephemeris values used in this analysis. We have used the NASA ephemeris calculations
\footnote{see \url{http://sohowww.nascom.nasa.gov/soc/mercury2003/}} provided by NASA for the SOHO Mercury transit observations.
Ephemeris uncertainties for the absolute contact times are not greater than 0.18 s based on the SOHO absolute position error.
Using the above equation we found the solar radius values to be
$960".03 \pm 0".08 $  $(696,277 \pm 58 \mathrm{km})$ and $960".07 \pm 0".05 $ $(696,306 \pm 36 \mathrm{km})$ for the
2003 and 2006 transits respectively. Uncertainties were determined from the scatter in the two FB settings
from each epoch. We note that the observed absolute
transit contact times are offset 8 s in 2003 and 5 s in 2006, both inside our $2\sigma$ error.

Our accurate determination of the limb transit points, compared with the ephemeris predictions, implies
that the orientation of the solar image is slightly rotated from the solar north pole lying along the y axis.
We find that the true solar north orientation is rotated $7' \pm 1'$
counterclockwise in 2003 and $3'.3\pm 0'.3$ arcmin in 2006. Table~\ref{observations} summarizes our results from this analysis.

\begin{figure*}
 \centering
 \plotone{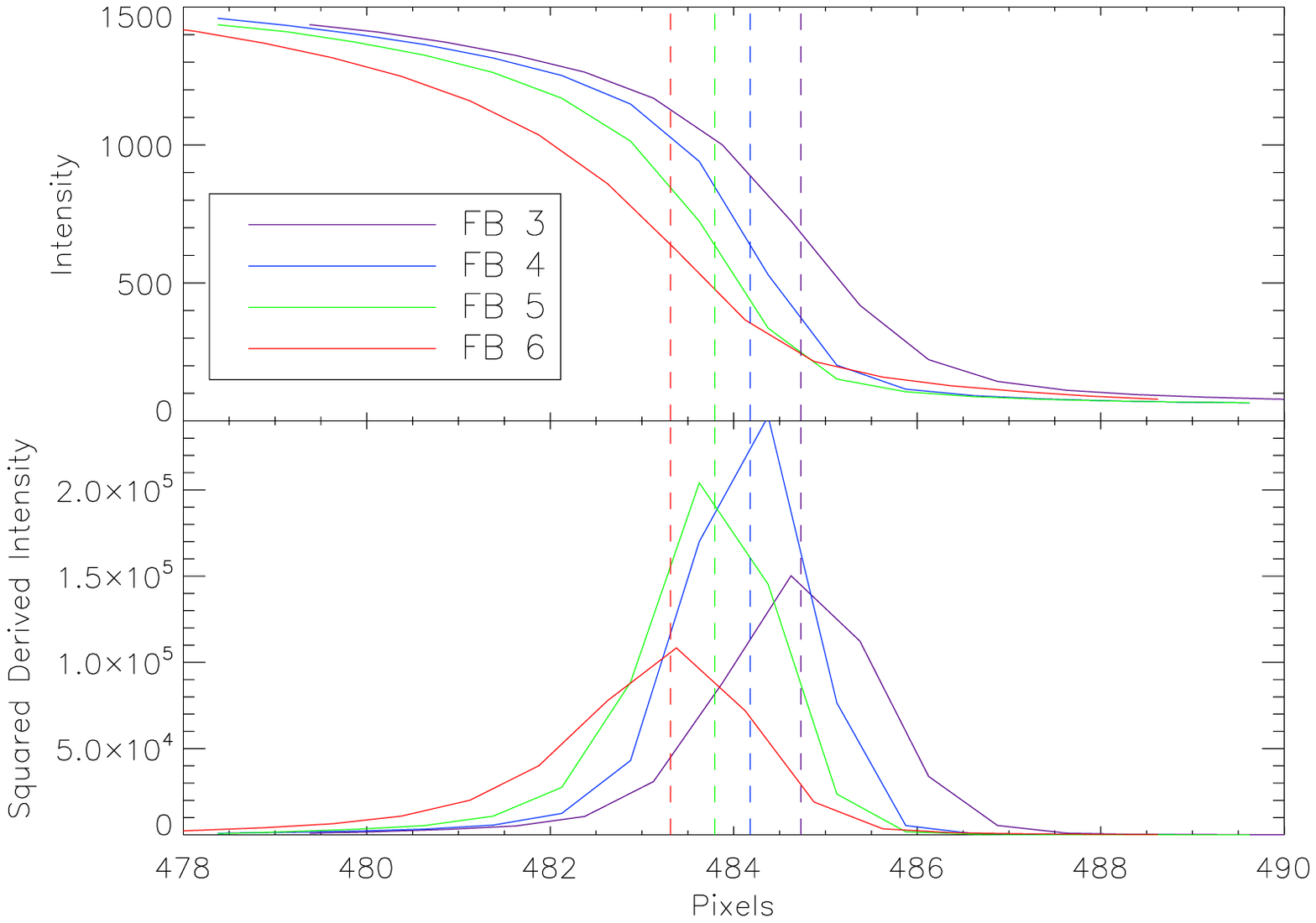}
 \caption{Top: mean LDfs at 2003 transit, one for each FB settings. Bottom: derivative squared of top Ldfs. The dashed lines indicate the inflection point location calculated from a Gaussian plus quadratic fit from the derivative squared points.}
 \label{ldfs}
\end{figure*}

\begin{deluxetable*}{ccccc}
\tabletypesize{\scriptsize}
\tablecolumns{5}
\tablecaption{Mercury transit observation}
\tablewidth{450pt}
\tablehead{
& \multicolumn{2}{c}{2003} & \multicolumn{2}{c}{2006}\\
\cline{2-5}  \\
Focus Settings & Transit Duration\tablenotemark{a} & Solar Radius & Transit Duration\tablenotemark{a} & Solar Radius \\
 & (minute) & (arcsec) & (minute) & (arcsec)
}
\startdata
3 & 324.16 & 960.00 & \nodata & \nodata \\
4 & 324.11 & 959.92 & 296.74 & 960.03\\
5 & 324.25 & 960.13 & 296.77 & 960.10\\
6 & 324.33 & 960.24 & \nodata & \nodata \\
\cline{2-5}  \\
Average\tablenotemark{b} & 324.18 & 960.03 & 296.75 & 960.07 \\
 \cline{2-5}  \\
p(O-C) & \multicolumn{2}{c}{$7' \pm 1'$} & \multicolumn{2}{c}{$3'.3 \pm 0'.3$}\\
\enddata
\tablenotetext{a}{Transit duration for Mercury transit contact times}
\tablenotetext{b}{Only FB 4 and FB 5 focus settings were used for the
2003 data (see the text for details).}
\label{observations}
\end{deluxetable*}

\begin{figure}
 \centering
 \plotone{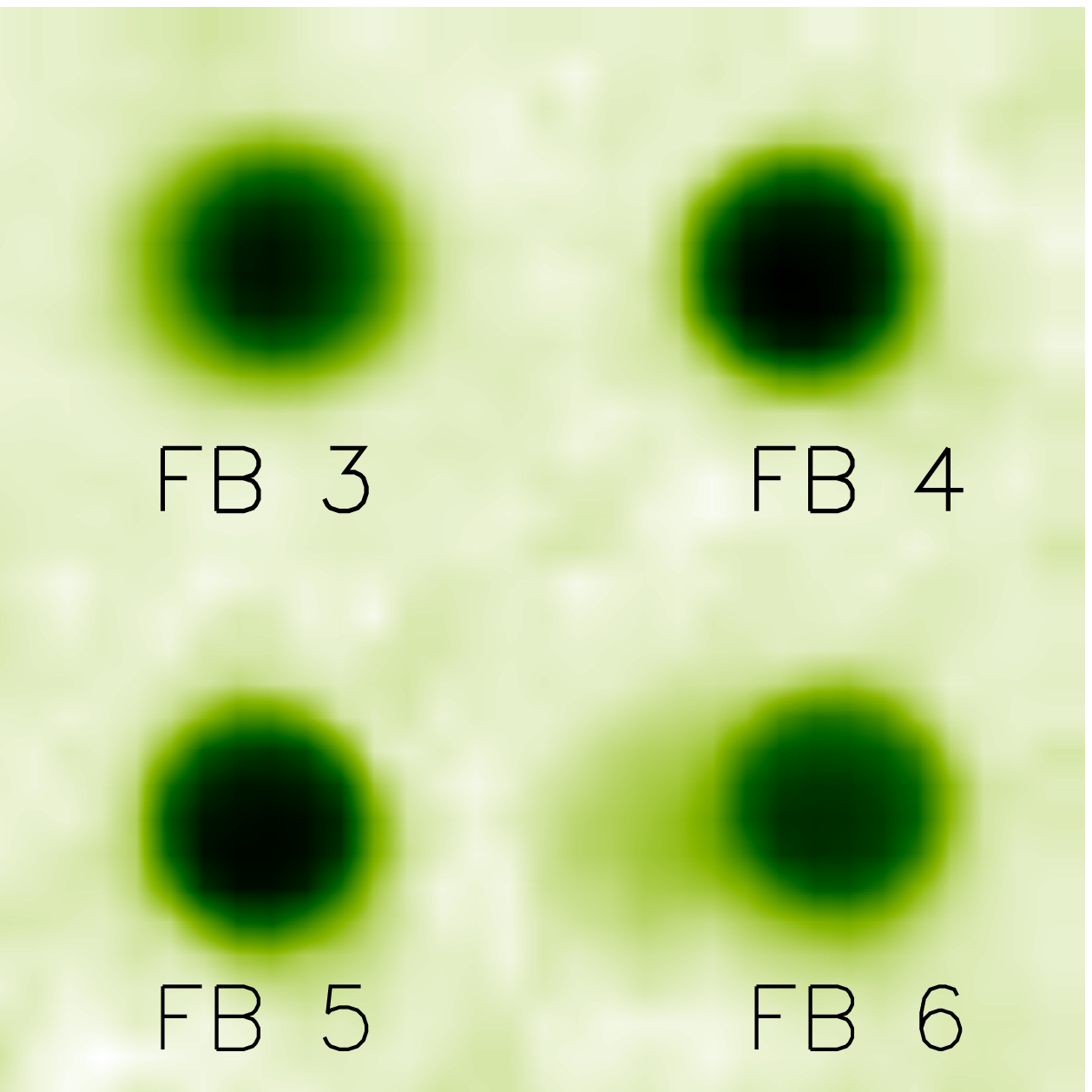}
 \caption{Close-up of Mercury observed by MDI. FB 6 is far from the instrument focal plane and has large systematics including a spurious ghost image.}
 \label{mercuryfb}
\end{figure}

\section{Systematics deviations}

\subsection {Numerical Limb Definition}
    The solar radius in theoretical models is defined as the photospheric region where optical depth is equal to the unity. In practice, helioseismic inversions determine this point
  using f mode analysis, but most experiments which measure the solar radius optically use the inflection point of the LDF as the definition of the solar radius. \citet{Tripathy99} argue that the differences between the two definitions are between 200 and 300 km (0".276-0".414). This explains why the two helioseismologic measurements in Figure~\ref{lradius} show a smaller value. The convolution of the LDF and Earth's atmospheric turbulence and transmission cause the inflection point to shift. The effect (corrected by the Fried parameter to infinity) is at the order of 0".123 for $r_0 = 5 $cm and 1".21 $r_0 = 1 $cm  \citep{djafer08} and depends on local atmospheric turbulence, the aperture of the instrument and the wavelength resulting in an observed solar radius smaller than the true one \citep{Chollet99}. Many of the published values were not corrected for the Fried
  parameter which may explain most of the low values seen in Figure~\ref{lradius}.

We analyzed some systematics regarding the numerical calculation of the inflection point. The numerical method to calculate the LDF derivative used is the five-point rule of the Lagrange polynomial. The systematic difference between the three-point rule that is used by default in the interactive data language derivative routine, is 0.01 pixels for FB 5 up to 0.04 pixels for FB 6. In this work we defined the inflection point as the maximum of the LDF derivative squared. From those points we fit a Gaussian plus quadratic function and adopt the maximum of this function as the inflection point. Adopting the maximum of the Gaussian part of the fit instead differs 0.001 pixels for FB 3,4,5 and five times more for FB 6. The quadratic function part is important since the LDF is not a symmetric function. Figure~\ref{ldf_time} compares the gaussian plus quadratic fit to find the function maximum with only fitting a parabola using 3 points near the maximum. They differ 0.01, -0.06, 0.04 and -.05 pixels for FB 3, 4, 5 and 6 respectively. The statistical error sampling for all the available data after accounting for changing the distance is 0.0002 pixels for FB 5 and 0.0003 pixels for FB 6. Figure~\ref{mercuryfb} shows a close up of Mercury for each focus block where we can visually see a spurious ghost image of FB 6 since this FB is  far from the focus plane. That can explain why the numerical methods gave more difference for this FB. By fitting a parabola over the solar image center positions as a function of time, we also took into account jitter movements of the SOHO spacecraft. Solar limb coordinates and the Mercury trajectory were adjusted to the corrected solar center coordinate frame. This procedure made a correction of order of 0.01 pixels in our definition of the solar limb. Smoothing the LDF derivative changes the inflection point up to .35 pixels that is the major correction we applied in our previous published value in \citet{kuhn04}. We describe all corrections applied to this value at the end of this section.

\subsection {Wavelength}
  MDI made solar observations in five different positions of a line center operating wavelength of 676.78 nm. Conforming Figure 1. of \citet{bush10} at line center, the apparent solar radius is 0.14 arcsec or approximately 100 km larger than the Sun observed in nearby continuum. We used a single filtergram nearby the continuum in this work. Comparison with the composite linear composition of filtergrams representing the continuum differs from 0.001 arcmin.  \citet{neckel95} suggest the wavelength correction due to the (continuum) increases with wavelength. Modern models confirm this dependence with a smaller correction. \citet{djafer08} computed the contribution of wavelength using \citet{HM98} model. According to this model and others cited by \citet{djafer08} the solar radius also increases with wavelength (see Fig. 1 of \citet{djafer08}). The correction at 550 nm is on the order of 0".02 and was not applied since the dependence of wavelength for the point-spread function (PSF) contribution goes to the opposite direction and the correction is inside our error bars.

\subsection {Point-spread-Function (PSF)}
   The MDI PSF is a complicated function since different points of the image are in different focus positions
(see \citet{kuhn04} for description of MDI optical distortions). In our earlier work the non-circular
low-order optical aberrations are effectively eliminated by averaging the radial limb position around the limb, over all angular bins.
But in the case of the Mercury transit, only two positions at the limb apply so that it is important to recognize that the LDF (and the
inflection point) will systematically vary around the limb. The overall contribution to the PSF of the solar image increases with increasing telescope aperture, and decreases with increasing observing wavelength as noted in the last section. \citet{djafer08} computed the contribution of the MDI instrument PSF at its center operating wavelength of 676.78 nm assuming a perfectly focused telescope dominated by diffraction.
The authors made use of the solar limb model of \citet{HM98} to calculate the inflection point. They found that the MDI instrumental LDF inflection point was displaced 0".422 inward
at `ideal focus' from the undistorted LDF, but the authors used a MDI aperture of 15 cm instead of actual aperture diameter of 12.5 cm. In addition, we do not find the pixelization effect cited by the authors.
PSF and wavelength correction for SDS calculated by \citep{djafer08} resulted in a diameter of $959".798 \pm 0".091$ \citep{djafer08}. The original value found by Calern astrolabe (CCD observations) is $959".44 \pm 0".02$ \citep{Chollet99}. After applying a model taking into account atmospheric and instrumental effects (with Fried parameter, $r \rightarrow \infty$) \citet{Chollet99} found $959".64 \pm 0".02$. From this value \citet{djafer08} applied PSF and wavelength correction and obtained $959".811 \pm 0.075$.

\subsection {Focus variation}

The overall focus of the instrument is adjusted by inserting BK7 and SF4 glass of different thickness into the optical path after the
secondary lens in order to vary the optical path length. The MDI focus changes occur in nine discrete increments using
sets of three glass blocks in the two calibration focus mechanisms. Increasing the FB number by one unit
moves the nominal focus toward the primary lens by 0.4 mm.
As described by \citet{kuhn04} a change in the focus condition causes a net change in the
image plate scale, but it also changes the shape of the LDF which affects the
inflection point. Figure~\ref{ldfs} plots the mean LDF averaged around the
limb for the various limb focus conditions during the two transits. The curves
are shifted to account for the plate scale change by matching the half-intensity
distance. The data here show how the defocused image causes the inflection
point to move to larger radii, thus implying a larger solar diameter by as much
as a few tenths of an arcsec for out-of-focus images.
The Mercury image will also be subject to the same distortion, but the inflection point moves from the first/second contact to
the third/forth contact. We see this systematic at different FB since the total transit time grows for bigger focus settings (see Table~\ref{observations}).

Figure~\ref{fbfit} (top) shows the best focus setting over the MDI history. From a polynomial fit we derived the best focus position
for the transit. For 2003, this value is 4.94 FB units (solid line) and in 2006 it was 5.35 (long dashed line).
We made a first order correction to account for the out-of-focus condition
of the instrument during each transit.
Figure~\ref{fbfit} bottom shows the values found for the solar radius at different focus settings. The 2003 transit values are represented
as "diamonds" and 2006 transits as "upper triangles". As we have only two focus settings in 2006 we represent the function as a line. As
the best focus is not so far from FB 5, the systematic error for adopting a line is small. The vertical lines (for 2003 we plot a short
dashed line and 2006 the long dashed line) represent the best focus position for each year. With a linear interpolation we found the
effective solar radius at best focus for both transits. The 2003 and 2006 focus correction resulted in a value of $960".12$ for both years.

\subsection {Systematics Effects from our previous 2004 MDI Solar Radius Determination}

The solar radius we determine from the transit data is bigger than our published
2004 MDI results that were obtained from a detailed optical distortion model. In the process of
understanding the transit measurements here we have found several systematic effects
that affect these and our 2004 measurements. Observing from space allows us to
measure and correct for field-dependent PSF errors that would normally be too
small to separate from an atmospheric seeing-distorted solar image.

The largest residual error comes from the variation
in focus condition across the image. Figure~\ref{ldfs} shows four typical LDF's at the middle of the 2003 transit (one for each FB).
Each LDF in Figure~\ref{ldfs} is evaluated as a mean over the whole image.
In contrast with \cite{kuhn04}, here we calculate mean LDFs around the
solar limb. This is accomplished by using the computed limbshift at each
angle (which depends on the average LDF) to obtain a distortion corrected
mean LDF. We iterate with this corrected LDF to obtain improved limb shift
data and an unsmeared LDF (see \citet{emilio07}). We have also improved the LDF inflection point
calculation from the local unsmeared LDF following the discussion in Section 3.1. Applying the above modifications the solar radius in pixels at the 2003 transit for FB 4 is 484.192 pixels at the middle of the transit (it was 484.371 in \citet{kuhn04}).

Our 2004 result obtained the mean image platescale by combining a detailed
optical raytrace model and the measured solar radius change with FB.
We also derived what we called the ``symmetric pincushion image distortion" from the  apparent path of Mercury across the solar image. These two effects are
contained in the empirical plate-scale measurement so that we should not have independently applied the symmetric distortion correction to the corrected plate-scale-data.
After correcting the plate-scale only a non-circularly symmetric correction (-0.32 pixels) was needed. With this correction we obtain $483.872 \pm 0.05$ pixels for the FB4 2003 Mercury transit solar radius. With the FB 4 calibration of $1.9870 \pm 0.0002$ pixel$^{-1}$ the corrected solar radius is $961".45 \pm 0".15$ (as seen from SOHO). During the middle of the 2003 transit, SOHO was 0.99876 AU from the Sun, so that
the apparent solar radius at 1 AU is then $960".26 \pm 0".15$. This value is consistent with the $1 \sigma$ uncertainty in the Mercury timing radius
derived here.

\begin{figure}
 \centering
 \plotone{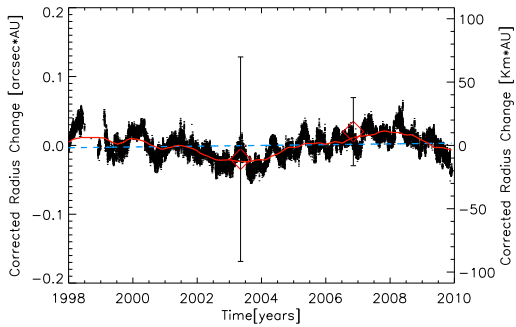}
 \caption{Residual variation of the solar radius (dots; \citet{bush10}) and the residual values found
 in this work for the Mercury transit after averaging the focus blocks settings ("diamonds"). Solid line represents the moving
 average values and the dashed line a linear fit from all the residual variation of the solar radius.}
 \label{variation}
\end{figure}

\section{Conclusion}

We now find a consistent value for the solar
radius from Mercury transit timing, and satellite solar imagery using MDI.  The averaged O-C transit duration uncertainties are 3 s
in 2003 and 1 s in 2006. The value we found for the
solar radius as defined by the apparent inflection point of the continuum
filtergram images of MDI is $960".12 \pm 0".09$ $(696,342 \pm 65 \mathrm{km})$.
This is consistent with earlier MDI absolute radius measurements after taking into account systematic corrections and a calibration error in the 2004
optical distortion measurements. We also obtain an MDI solar reference frame,
counterclockwise, p angle correction of $7' \pm 1'$
in 2003 and $3'.3 \pm 0'.3$ in 2006. Figure~\ref{variation} shows the residual variation of the solar radius \citep{bush10} and the residual values found
 in this work for the Mercury transit after averaging the focus blocks settings. As seen in Figure~\ref{variation} no variation of the solar radius was observed over 3 years between the two transits.

\acknowledgments
We are grateful to Manuel Montoro from the Flight Dynamics Facility Goddard Space Flight Center to provide us with updated ephemeris.
This research was supported by the NASA SOHO/GI program through a grant to the IfA, by the NASA Michelson Doppler Imager grant NNX09AI90G to Stanford and the NASA Helioseismic and Magnetic Imager contract NAS5-02139 to Stanford and subcontract to the IfA. This work was also partially supported by CNPq grant 303873/2010-8, the Instituto Nacional de Estudos do Espa\c{c}o (CNPq) and CAPES grant 0873/11-0.

\end{document}